\documentclass[twocolumn,notitlepage,]{article}
\def\1{{\bf 1}}

\def\Tr{{\rm Tr}}
\def\C{{\bf C}}
\def\R{{\bf R}}
\def\E{{\cal E}}

\def\mfr#1/#2{\hbox{${{#1} \over {#2}}$}}
\def\const.{{\rm const.}}
\def\beq{\begin{equation}}
\def\eeq{\end{equation}}
\def\beqa{\begin{eqnarray}}
\def\eeqa{\end{eqnarray}}
\begin{document}
\setlength{\unitlength}{1.0cm}
%\title{\bf \\
%A rigorous derivation
%\footnotetext{\copyright 1998 by Elliott H. Lieb}

\def\boxit#1{\thinspace\hbox{\vrule\vtop{\vbox{\hrule\kern1pt
\hbox{\vphantom{\tt/}\thinspace{\tt#1}\thinspace}}\kern1pt\hrule}\vrule}
\thinspace}

\font\eightit=cmti8
\def\ve{{\varepsilon}}
\def\C{{\bf C}}
\def\D{{\rm D}}
\def\E{{\rm E}}
\def\R{{\bf R}}
\def\Z{{\bf Z}}
\def\L{{\Lambda}}
\def\l{{\lambda}}
\def\d{{\rm d}}
\def\Q{{{\rm C}_{\Lambda}}}
\def\X{{\underline{X}}}
\def\Tr{{\rm Tr}}
%%%%%%%%%%%%%%%%%%%%%%%%%%%%

%%%%%%%%%%%%%%%%%%%%%%%%%%%%

{\bf LIEB-THIRRING INEQUALITIES} -- Inequalities concerning the
negative eigenvalues of the {\bf Schr\"odinger operator}
$$
H = - \Delta + V(x)
$$
on $L^2(\R^n), n \geq 1$. With $e_1 \leq e_2 \leq \cdots < 0$
denoting the negative eigenvalues of $H$ (if any), the Lieb-Thirring
inequalities 
state that for suitable $\gamma \geq 0$ and constants $L_{\gamma,n}$
\begin{equation}
\sum_{j \geq 1} |e_j|^{\gamma} \leq L_{\gamma,n} \int_{\R^n} V\_
(x)^{\gamma + n/2} \ \d x 
\end{equation}
with $V\_ (x) := {\rm{max}} \{ - V (x),0 \}$. When $\gamma = 0$ the
left side is just the number of negative eigenvalues. Such an
inequality (1) can hold if and only if
\begin{eqnarray}
\gamma \geq \frac{1}{2} & {\rm{for}} & {\rm{n}} = 1 \nonumber \\
\gamma > 0 & {\rm{for}} & {\rm{n}} = 2\\
\gamma \geq 0 & {\rm{for}} & {\rm{n}} \geq 3 \ .  \nonumber 
\end{eqnarray}

The cases $\gamma > \frac{1}{2}, n = 1, \gamma > 0, n \geq 2$, were
established by E.H. Lieb and W.E. Thirring~\cite{LT} in connection with
their proof of {\bf stability of matter}. The case $\gamma =
\frac{1}{2}, n = 1$ was established by T. Weidl~\cite{TW}. The case
$\gamma = 0$, $n \geq 3$ was established independently by M.
Cwikel~\cite{MC}, Lieb~\cite{EL} and G.V. Rosenbljum~\cite{GR} by
different methods and is known as the CLR bound; the smallest known
value for $L_{0,n}$ is in~\cite{EL},~\cite{EL2}.

Closely associated with the inequality (1) is the {\it semi-classical
approximation} for $\sum|e|^{\gamma}$, which serves as a heuristic
motivation for (1). It is ({\it cf.}~\cite{LT}).
\begin{eqnarray*}
\sum_{j\geq1} |e|^{\gamma} &\approx& (2 \pi)^{-n} \int_{\R^n \times
\R^n} \left[ p^2 + V(x) \right]^{\gamma}_{\_} \d p \d x\\
& =& L_{\gamma,n}^{c} \int_{\R^n} V \_ (x)^{\gamma + n/2} \ \d x
\end{eqnarray*}
with 
$$
L_{\gamma,n}^{c} = 2^{-n} \pi^{-n/2} \Gamma (\gamma + 1)/\Gamma
(\gamma + 1 + n/2) \ .
$$
Indeed, $L_{\gamma, n}^{c} < \infty$ for all $\gamma \geq 0$ whereas
(1) holds only for the range given in (2). It is easy to prove (by
considering $V(x) = \lambda W (x)$ with $W$ smooth and $\lambda
\rightarrow \infty$) that
$$
L_{\gamma, n} \geq L_{\gamma, n}^{c}
$$

An interesting, and mostly open problem is to determine the sharp value
of the constant $L_{\gamma, n}$, especially to find those cases in which
$L_{\gamma,n} = L_{\gamma, n}^{c}$. M. Aizenman and Lieb~\cite{AL}
proved that the ratio $R_{\gamma,n} = L_{\gamma,n} / L_{\gamma,
n}^{c}$ is a monotonically non-increasing function of $\gamma$. Thus,
if $R_{\Gamma,n} =1$ for some $\Gamma$ then $L_{\gamma,n} =
L_{\gamma,n}^{c}$ for all $\gamma \geq \Gamma$. The equality
$L_{\frac{3}{2},n} = L_{\frac{3}{2},n}^c$ was proved for $n=1$ in 
~\cite{LT} and for $n>1$ in \cite{LW} by A. Laptev and T. Weidl. (See
also \cite{BL}.)

The following sharp constants are known: 
\begin{eqnarray*}
L_{\gamma,n} &=& L_{\gamma,n}^{c} \qquad  {\rm{all}}~ \gamma \geq
3/2,~\cite{LT},~\cite{AL},~\cite{LW} \\ 
L_{1/2, 1} &=& 1/2 \qquad \qquad \qquad \cite{HLT}\\
\end{eqnarray*}

There is strong support for the conjecture~\cite{LT} that
\begin{equation}
L_{\gamma,1} = \frac{1}{\sqrt{\pi} (\gamma - \frac{1}{2})} \frac{\Gamma
( \gamma + 1)}{\Gamma (\gamma + 1/2)} \left( \frac{\gamma -
\frac{1}{2}}{\gamma + \frac{1}{2}} \right)^{\gamma + 1/2}
\end{equation}
for $\frac{1}{2} < \gamma < \frac{3}{2}$.

Instead of considering all the negative eigenvalues as in (1), one can
consider just $e_1$. Then for $\gamma$ as in (2)
$$
|e_1|^{\gamma} \leq L_{\gamma, n}^1 ~ \int_{\R^n} V\_(x)^{\gamma +
 n/2} \d x \ .
$$
Clearly, $L_{\gamma, n}^1 \leq L_{\gamma, n}$, but equality can hold,
as in the cases $\gamma = 1/2$ and $ 3/2$ for $ n = 1$. Indeed, the
conjecture in (3) amounts to $L_{\gamma,1}^1 = L_{\gamma, 1}$ for $1/2
< \gamma < 3/2$. The sharp value (3) of $L_{\gamma, n}^1$ is obtained by
solving a differential equation~\cite{LT}. It has been conjectured that
for $n \geq 3, L_{0, n} = L_{0, n}^1$. In any case, B. Helffer and D.
Robert showed that for all $n$ and all $\gamma < 1$, $L_{\gamma, n} >
L_{\gamma, n}^c$.

The sharp
constant $L_{0, n}^1, n \geq 3$ is related to the sharp constant $S_n$
in the {\bf Sobolev inequality}
\begin{equation}
\parallel \nabla f \parallel_{L^2(\R^n)} \geq S_n \parallel ~ f ~
\parallel_{L^{2n/(n-2)}(\R^n)}
\end{equation}
by $L_{0, n}^1 = (S_n)^{-n}$.

By a `duality argument'~\cite{LT} the case $\gamma =1$ in (1) can be converted
into the following bound for the Laplacian, $\Delta$. This bound is
referred to as a Lieb-Thirring kinetic
energy inequality and its most important application is to the {\bf
stability of matter}~\cite{EL3}, \cite{LT}.
Let $f_1, f_2, \ldots$ be {\it any} orthonormal sequence (finite or
infinite) in $L^2 (\R^n)$ such that $\nabla f_j \in L^2 (\R^n)$ for
all $j \geq 1$. Associated with this sequence is a `density'
\begin{equation}
\rho(x) = \sum_{j \geq 1} | f_j (x) |^2 \ .
\end{equation}
Then, with $K_n := n(2/L_{1, n})^{2/n} (n+2)^{-1-2/n} \ ,$
\begin{equation}
\sum_{j \geq 1} \int_{\R^n} | \nabla f_j (x)|^2 \d x \geq K_n
\int_{\R^n} \rho(x)^{1 + 2/n} \d x \ .
\end{equation}
This can be extended to {\it antisymmetric} functions in $L^2
(\R^{nN})$. If $\Phi = \Phi (x_1, \ldots , x_N)$ is such a function we
define, for $x \in \R^n$,
$$
\rho(x) = N \int_{\R^{n(N-1)}} | \Phi (x, x_2, \ldots , x_N)|^2 \d x_2
\ldots \d x_N \ .
$$
Then, if $\int_{\R^{nN}} | \Phi |^2 = 1$,
\begin{equation}
\int_{R^{nN}} | \nabla \Phi|^2 \geq K_n \int_{\R^n} \rho (x)^{1 + 2/n}
\d x \ .
\end{equation}
Note that the choice $\Phi = (N!)^{-1/2} \det f_j(x_k)|^{N}_{j, k=1}$
with $f_j$ orthonormal reduces the general case (7) to (6).

If the conjecture $L_{1, 3} = L_{1, 3}^c$ is correct then the bound in
(7) equals the {\bf Thomas-Fermi} kinetic energy {\it  ansatz}, and
hence it is a challenge to prove this conjecture. In the meantime, 
see ~\cite{EL2}, ~\cite{BS} for
the best available constants to date (1998).

Of course, $\int (\nabla f)^2 = \int f (-\Delta f)$. Inequalities of
the type (7) can be found for other powers of $-\Delta$ than the first
power. The first example of this kind, due to I. Daubechies~\cite{ID},
and one of the most important physically, is to replace $-\Delta$ by
$\sqrt{-\Delta}$ in $H$. Then an inequality similar to (1) holds with
$\gamma + n/2$ replaced by $\gamma + n$ (and with a different
$L_{\gamma, n_{1}}$, of course). Likewise there is an analogue of (7) with
$ 1 + 2/n$ replaced by $1 + 1/n$.

All proofs of (1) (except \cite{HLT} and \cite{TW})actually proceed by
finding an upper bound to $N_E (V)$, the number of eigenvalues of $H =
-\Delta + V(x)$ that are below
$-E$. Then, for $\gamma > 0$, $$ \sum |e|^{\gamma} = \gamma
\int_{0}^{\infty} N_E (V) E^{\gamma - 1} \d E
$$
Assuming $V = -V\_$ (since $V_+$ only raises the eigenvalues),
$N_E(V)$ is most accessible via the positive semidefiniate {\it
Birman-Schwinger kernel} ({\it cf.} \cite{BSi})
$$
K_E (V) = \sqrt{V\_}~ ( -\Delta + E)^{-1} \sqrt{V\_} \ .
$$
$e < 0 $ is an eigenvalue of $H$ if and only if 1 is an eigenvalue of
$K_{|e|}(V)$. Furthermore, $K_E(V)$ is {\it operator}  monotone
decreasing in $E$, and hence $N_E(V)$ equals the
number of eigenvalue of $K_E(V)$ that are greater than 1.

An important generalization of (1) is to replace $- \Delta$ in $H$ by
$|i \nabla + A (x)|^2$, where $A(x)$ is some arbitrary vector field in
$\R^n$ (called a magnetic vector potential). Then (1) still holds but
it is not known if the sharp value of $L_{\gamma, n}$ changes. What is
known is that all {\it presently} known values of $L_{\gamma, n}$ are
unchanged. It is also known that $(- \Delta + E)^{-1}$, as a kernel in
$\R^n \times \R^n$, is pointwise greater than the absolute value of
the kernel $(| i \nabla + A |^2 + E)^{-1}$.

There is another family of inequalities for orthonormal functions,
which is closely related to (1) and to the CLR bound~\cite{EL4}. As
before, let $f_1, f_2, \ldots , f_N$ be $N$ orthonormal functions in
$L^2(\R^n)$ and set
\begin{eqnarray*}
u_j &=& ( - \Delta + m^2)^{-1/2} f_j\\
\rho (x) &=& \sum_{j = 1}^{N} |u_j (x)|^2 \ .
\end{eqnarray*}
$u_j$ is a {\bf Riesz potential} ($m=0$) or {\bf Bessel potential} ($m
> 0$) of $f_j$. 
If $n = 1$ and $m > 0$ then, $\rho \in C^{0, 1/2}(\R^n)$ and
$\parallel \rho \parallel_{L^{\infty}(\R)} \leq L/m$.

If $n = 2$ and $m > 0$ then for all $1 \leq p < \infty \\
\noindent \parallel\rho
\parallel_{L^p(\R^2)} \leq B_p m^{-2/p} N^{1/p}.$

If $n \geq 3, p = n/(n-2)$ and $m \geq 0$ (including $m = 0$) then
$\parallel \rho \parallel_{L^p(\R^n)} \leq A_n N^{1/p}.$

Here, $L, B_p, A_n$ are universal constants. Without the orthogonality,
$N^{1/p}$ would have to be replaced by $N$. Further generalizations
are possible~\cite{EL4}.

\bigskip

\noindent \hfill{\it Elliott H. Lieb}\\
\noindent \rightline{\it Departments of Mathematics and Physics}
\noindent \rightline{\it Princeton University}\\

\bigskip
\footnoterule
\bigskip
\noindent {\copyright 1998 by Elliott H. Lieb}

\end{document}